\documentclass[floatfix,twocolumn,aps,superscriptaddress,longbibliography,10]{revtex4-1}
\usepackage{rotating}
\usepackage{graphicx}
\usepackage{bm}

\begin{document}

\title{Spreader events and the limitations of projected networks for capturing dynamics on multipartite networks}

\author{Hyojun A. Lee}
\thanks{These authors contributed equally to this work}
\affiliation{Department of Chemical and Biological Engineering, Northwestern University, Evanston, Illinois, United States of America}

\author{Luiz G. A. Alves}
\thanks{These authors contributed equally to this work}
\affiliation{Department of Chemical and Biological Engineering, Northwestern University, Evanston, Illinois, United States of America}%

\author{Lu\'is A. Nunes Amaral}
\email{amaral@northwestern.edu}
\affiliation{Department of Chemical and Biological Engineering, Northwestern University, Evanston, Illinois, United States of America}
\affiliation{Department of Physics and Astronomy, Northwestern University, Evanston, Illinois, United States of America}
\affiliation{Northwestern Institute on Complex Systems, Northwestern University, Evanston, Illinois, United States of America}

\date{\today}

\begin{abstract}
Many systems of scientific interest can be conceptualized as multipartite networks. Examples include the spread of sexually transmitted infections, scientific collaborations, human friendships, product recommendation systems, and metabolic networks. In practice, these systems are often studied after projection onto a single class of nodes, losing crucial information. Here, we address a significant knowledge gap by comparing transmission dynamics on temporal multipartite networks and on their time-aggregated unipartite projections to determine the impact of the lost information on our ability to predict the systems' dynamics. We show that the dynamics of transmission models can be dramatically dissimilar on multipartite networks and on their projections at three levels: final outcome, the magnitude of the variability from realization to realization, and overall shape of the temporal trajectory. We find that the ratio of the number of nodes to the number of active edges over the time aggregation scale determines the ability of projected networks to capture the dynamics on the multipartite network. Finally, we explore which properties of a multipartite network are crucial in generating synthetic networks that better reproduce the dynamical behavior observed in real multipartite networks. 
\end{abstract}

\maketitle
\section*{Introduction}

The theoretical study of transmission processes, whether infectious diseases, innovations, or mores, has typically followed one of two approaches: compartment models~\cite{daley2001epidemic} and network models~\cite{pastor2001epidemic,Liljeros2003,barrat2008dynamical}. In compartment models, individuals belong to a small number of compartments, and they transit in and out of compartments according to specific rates. Compartment models, are mean-field models in which all members of a compartment are equally likely to interact with members of another compartment. In network models, individuals interact pairwise according to the connections defined in the underlying social network. Both approaches have produced significant advances both separately and when integrated as can be seen from the studies and forecasting of the current SARS-CoV-2 pandemic and of other recent pandemics~\cite{chinazzi2020effect,maier2020effective}.

Research published in the past two decades has demonstrated that network representations provide a fruitful abstraction of many complex systems~\cite{Amaral2004,Newman2010}.  Despite the remarkable advances, the task of determining the simplest network representation able to fully capture the characteristics of the system remains more of an art than a science. Indeed, although limitations in data collection procedures and less developed theoretical frameworks lead many scholars to settle for static, unipartite, unweighted, and undirected network representations, recent research has demonstrated that such representations may be inadequate~\cite{Barrat2004,Stouffer2005,Buldyrev2010,Kivela2014}.

For example, in the context of human mobility or international trade, one cannot disregard the weights (i.e., number of passengers in transportation networks, or exchange volume in international trade flows) of the connections between system components~\cite{fagiolo2008topological,Alves2020}.  In cases such as these, the system is more accurately conceptualized through {\it weighted\/} networks~\cite{Barrat2004}.  Similarly, in the context, for example, of the spread of information within social systems, one cannot ignore the multiple media --- email, Twitter, WhatsApp, and so on -- that are used by individuals to communicate. In cases such as these, the system is more accurately conceptualized through {\it multilayer\/} networks~\cite{Buldyrev2010}.  As a result, in recent years there has been an important and thriving research stream focused on multiplexed systems~\cite{Kivela2014,Boccaletti2014}.

Another significant characteristic of many non-equilibrium complex systems is that interactions among components are not immutable~\cite{Gorochowski2017,Li2017,Miritello2011,Liu2014}.  For example, direct flight connections to a city can be added or dropped over time; friendships flourish and wane; trophic relations in an ecosystem change with the seasons. In cases such as these, the system is more accurately conceptualized through {\it temporal\/} networks \cite{Buldyrev2010,Holme2012}. The temporal aggregation of evolving networks can significantly affect processes such as transmission phenomena~\cite{Holme2012,Pastor2015}. As a result, a significant research thrust has focused on the study of temporal networks and its implications to dynamical processes~\cite{Boccaletti2006,Perra2012,takaguchi2013bursty,karimi2013threshold,perotti2014temporal,holme2018probing,Petri2018,DeDomenico2016,Krings2012,Clauset2012,Ribeiro2013,Darst2016}.

In contrast to the deserved attention provided by these important and flourishing research streams to weighted, multilayer and temporal networks, multipartite networks remain overlooked (see Ref.~\cite{Benson2018} for exceptions and Refs.~\cite{bisanzio2010modeling,wen2012global,han2018epidemic,pavlopoulos2018bipartite,cao2011rendezvous} for epidemics on bipartite networks). However, many important biological and social systems are most naturally conceptualized as multipartite networks. Examples include cellular metabolism~\cite{Maslov2002}, ecosystems~\cite{Pilosof2017}, collaborations~\cite{Borner2004}, heterosexual romantic relationships~\cite{Liljeros2003,Helleringer2007,Rocha2011}, research collaborations~\cite{Guimera2005a}, or product recommendation systems~\cite{Leskovec2006,Zhou2007,Godoy-Lorite2016}. These systems comprise multiple classes of components --- for example, movies, actors, directors, producers, agents, studios, and so on in movie-production networks --- and edges can only connect nodes of different types --- producers connect through movies or through studios. 

Despite the diversity of ways complex systems can be represented and projected onto complex networks, our understanding of how modeling choices affect the dynamical process remains limited. Presumably because of the difficulty in obtaining detailed data, or for simplicity of analysis~\cite{Rivera2010}. In particular,  multipartite networks are typically studied after projection onto a single type of node considered to be the most important, such as actors in movie-production networks. However, by construction, the unipartite projection of a multipartite network is dramatically less information rich than the real network. For this reason, there is growing interest and effort in the community to collect open-access, large-scale, and information rich datasets on different network projections with time dependence to better understand real networks (see, for instance, Ref.~\cite{datasets}).

\begin{figure*}[t]
\centerline{\includegraphics[width=0.8\textwidth]{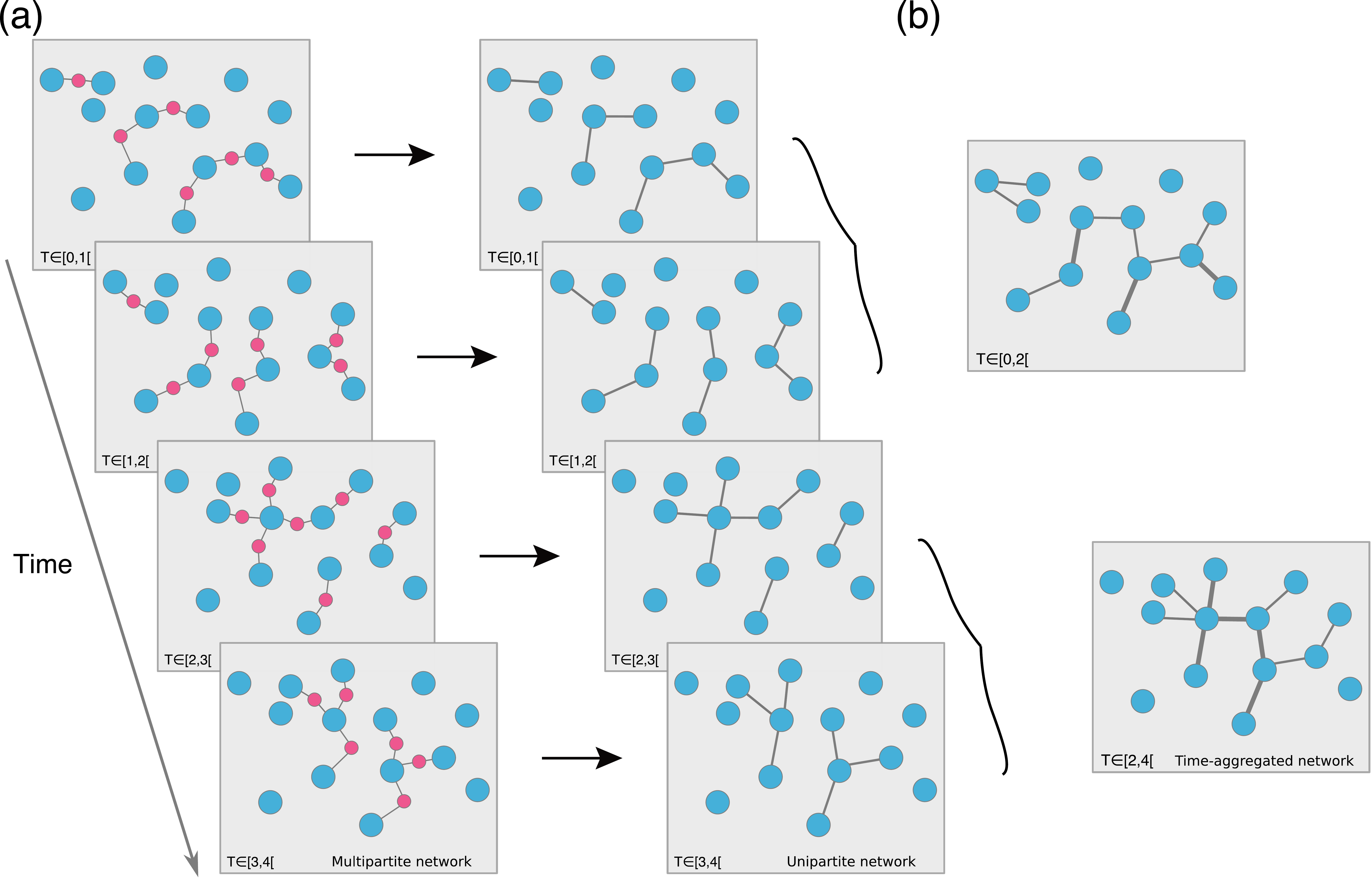}}
\caption{\textbf{Temporal multipartite network, unipartite projections, and time-aggregated networks.} (a) Temporal multipartite network (left) comprised of a class of agents nodes (larger blue circles) and a class of event nodes (pink circles) and its temporal agent projection (right). Event nodes have specific creation times and finite temporal duration (such as a day or a week). (b) Time-aggregated unipartite projection over the agent nodes shows simpler structures and dynamics. When the network is projected onto the agent network, nodes with temporal properties can be aggregated over different periods (a week, a month, and so on), losing crucial information on the ordering of events.}
\label{fig:fig1}
\end{figure*}

A particularly important aspect of some multipartite networks is the presence of strong temporal discontinuities for, at least, one type of node. These nodes present discrete timestamps and confined temporal existence: Movies have production schedules and romantic relationships span finite periods [Fig.~\ref{fig:fig1}(a)]. For this reason, these temporal multipartite networks have a stronger and more complex temporal structure than the typical temporal network and detailed temporal information is frequently unavailable [Fig.~\ref{fig:fig1}(b)].  Thus, it remains unclear whether the approach of temporal fine-graining used in the context of temporal networks to obtain static snapshots~\cite{Holme2005} would be sufficient to generate an accurate conceptualization of multipartite networks that include a class of nodes with heterogeneous and discontinuous event dynamics. 

To address this question, we investigate the extent to which unipartite projections of a multipartite network can be used to accurately describe and predict transmission dynamics~\cite{Kivela2014,Shakarian2015,Myers2012,Nematzadeh2014,Masuda2017a} taking place within multipartite networks in which one class of nodes has discontinuous dynamics.  As we mentioned earlier, we focus on transmission dynamics because they have been used to model a broad class of important phenomena such as epidemic outbreaks, opinion formation, and the spread of innovations~\cite{Darst2016,Valdano2015,Valdano2015a}.  Because of the broad range of applications of interest, we will consider model predictions for the full range of infectability values from the low infection rate of viruses, such as HIV~\cite{Rocha2011} to the high infection rate of ``unavoidable" memes~\cite{Gleeson2016}.

We organize our analysis around two systems with very different sizes and characteristics --- but for which we have high-quality data --- and that have been studied in the context of the adoption of innovations among small teams of highly trained professions~\cite{Weiss2014}, and in the context of the assembly of creative teams~\cite{Wasserman2015}.  We simulate transmission dynamics on these networks and on synthetic networks derived from them using two well-known transmission models --- susceptible infectious (SI) and general contagion (GC)~\cite{Anderson1992,Dodds2004,Dodds2005}. 

Significantly, we find under real-world conditions that the time-aggregated unipartite projections are broadly unable to predict the dynamics observed on the full temporal multipartite networks. Using model networks, we show that when the ratio of number of agents to the number of events/teams is less than one, unipartite projections will fail to reproduce the final number of infected agents, the magnitude of the variability from different realizations, and the overall shape of the temporal trajectories observed on multipartite networks. Furthermore, to address the cases in which unipartite projections are inadequate, we investigate the minimal amount of information that must be retained from a multipartite network in order to generate synthetic multipartite networks that properly reproduce the dynamics observed in the real networks.

\section*{Empirical multipartite networks and their projections} 

A multipartite network is a network whose nodes are partitioned into $K$ distinct non-overlapping sets. The nodes in each set have edges can connect only to nodes in the other $K-1$ sets [Fig.~\ref{fig:fig1}(a)]. We consider two contexts where connectivity is well described by multipartite networks, physician coverage teams in a hospital and movie-producing teams. 

For the former, we consider the multipartite networks formed by the physicians providing high-intensity critical care coverage in the medical intensive care unit (MICU) of a Chicago hospital that we studied previously~\cite{Weiss2014}. Specifically, we obtained the MICU coverage schedules of 76 physicians from July 2012 to August 2016. Each team is composed of one attending and one fellow and has confined temporal existence (see the Supplemental Material~\cite{SM} and Ref.~\cite{Weiss2014} for details). We construct a temporal multipartite network for each academic year comprised of three sets of nodes: coverage teams, attendings, and fellows. Attendings and fellows are only connected to team nodes. Edges between nodes are created when a fellow and an attending are together in the same team. Since the team has finite temporal existence, the edges exist only while the teams are active, typically a week.

For the latter, we construct a network of producer collaborations on U.S. movies, that we obtained by crawling the Internet Movie Database (IMDb, \url{https://imdb.com}). We focus here on production teams for the 2,009 US-produced movies released in the period 1990-2000. We identify 5,758 active producers during the considered period. Unlike the physician teams, we lack information on the exact production period for each movie.

Because the temporal aggregation time scale is known to affect static network structure~\cite{Krings2012,Holme2012,Pastor2015}, we compare the temporal multipartite physician network with different unipartite projected networks aggregated over different periods of time, spanning from 7 days to 12 months.  In the unipartite projected networks, the edges are static during the respective time period and therefore lose information about the temporal ordering of interactions~\cite{Zhou2007}. In addition to the unweighted unipartite network projection, we consider two weighting approaches for defining edge weight: the number of shared teams and the number of days in shared teams. For the movie producers, we aggregate time over four time scales 1, 2, 5, and 10 years. We model the edges as unweighted for the producer projected network because the vast majority of producers participate in only one movie per year.

\section*{Transmission dynamics}

Here, we focus on transmission dynamics and use the language of infectious diseases to investigate the limitations of time-aggregated unipartite projected networks for capturing the dynamics of temporal multipartite networks. The physicians' network is a tripartite network composed of two classes of agent nodes (attendings and fellows) and one class of event nodes (teams). At time zero, two agent-nodes are randomly chosen to become infectious (for the multipartite network, we chose the two nodes that are connected to the first event node), whereas all other nodes are deemed susceptible and to have no prior exposure (see Methods for details). We then track the number of infectious individuals over time.

\begin{figure*}[!t]
\centerline{\includegraphics[width=1\textwidth]{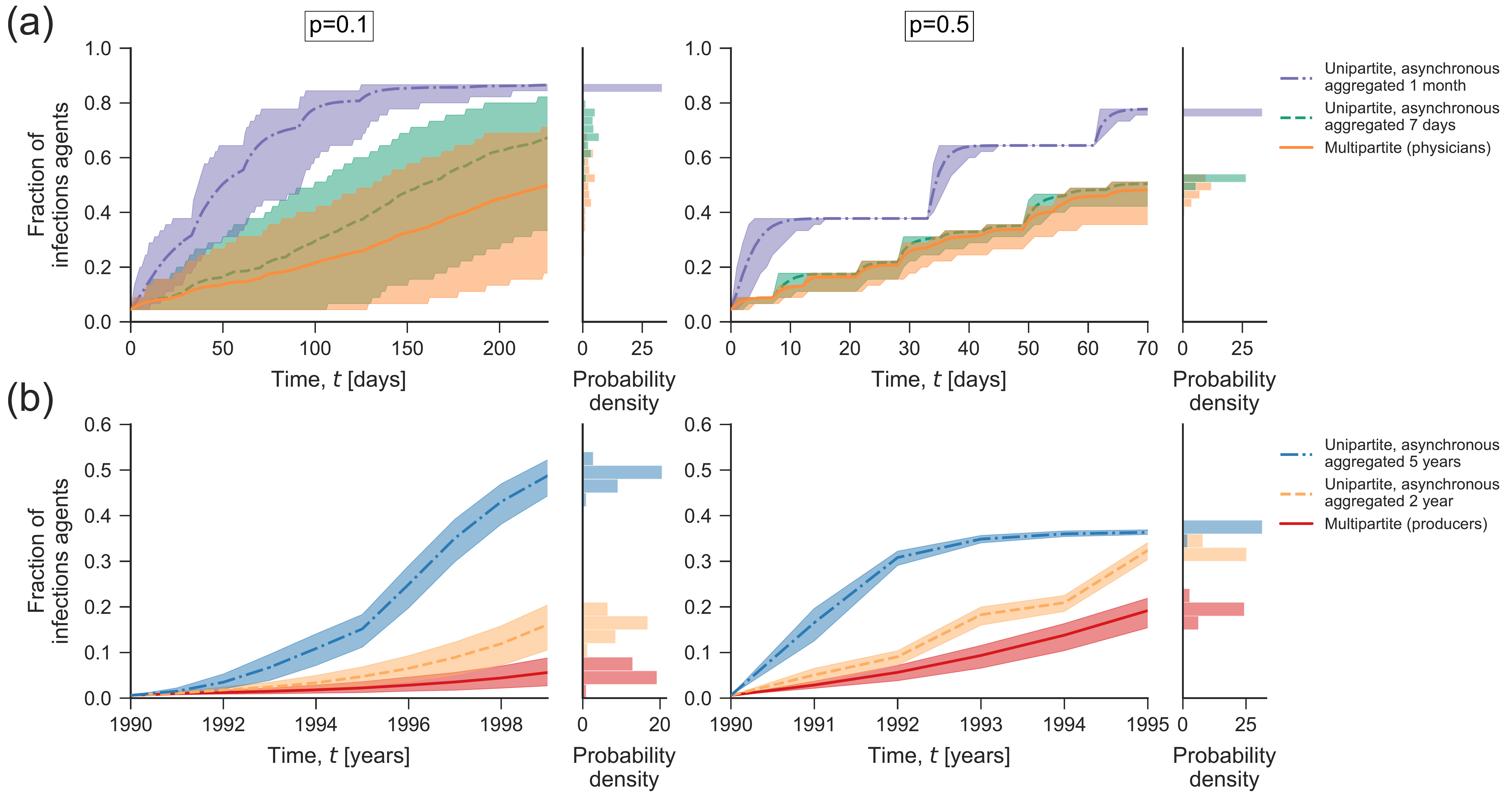}}
\caption{\textbf{Spreading dynamics on temporal multipartite networks and on their time-aggregated unipartite projections.} (a) Mean fractions of infectious agents (full lines) and 95\% confidence interval for the dynamics (shaded regions) on the physicians' multipartite networks and unweighted unipartite projected networks using 7 days and 1 month temporal aggregation (initial infectious agents: 2). The results for the projected network are obtained for the asynchronous status update. (b) Mean fractions of infectious agents (full lines) and 95\% confidence interval for the dynamics (shaded regions) on the producers' multipartite networks and unweighted unipartite projected networks using 2 and 5 year temporal aggregation (initial infectious agents: 5\% of producers appearing in the first year). The results for projected networks are obtained for the asynchronous status update. The right side of each panel shows the distributions of the final fraction of infectious agents for each set of parameters and conditions.}
\label{fig:fig2}
\end{figure*}

Figure~\ref{fig:fig2}(a) shows the mean number of infectious agents over time for two different parameter values for the SI model on multipartite and projected networks. It is visually apparent that the dynamics on the projected networks yield different dynamics from those observed on the multipartite network [Fig.~\ref{fig:fig2}(a)]. 

We further investigate the dynamic differences using the GC model to compare the temporal trajectories of the four multipartite networks.  Our benchmark multipartite networks (i.e., the networks for the different academic years) show similar temporal trajectories for a wide range of parameters, $0.1<d<1$ and $0.1<p<1$; Fig.~S1 of the Supplemental Material~\cite{SM}. In contrast, we can observe that the dynamics of time-aggregated unipartite projected networks compared with our benchmark multipartite network can overestimate the number of infectious agents over the course of time for a wide range of parameters for asynchronous dynamics simulations (see Figs.~S2 to S4 in the Supplemental Material~\cite{SM}), and it can overestimate or underestimate the number of infected agents depending on the time-aggregation interval and parameter range for synchronous dynamics simulations (see Figs.~S5 to S7 in the Supplemental Material~\cite{SM} for synchronous dynamics). We will provide a quantification of these differences in the following sections. 

Transmission dynamics on projected networks also lack the trajectory ``diversity" displayed by the dynamics on real networks [compare 95\% confidence intervals in Fig.~\ref{fig:fig2}(a)], pointing at deep differences of the dynamics.

The producers' network is a bipartite network where there are one class of agent nodes (producers) and one class of event nodes (movies). Using the SI model to simulate the transmission process on temporal multipartite networks and compare with their projections, we find that the movie producers network exhibit less variable temporal trajectories than the physicians' networks [Fig.~\ref{fig:fig2}(b)]. Nonetheless, it is visually apparent that the dynamics on the projected movie-producers' networks yield different dynamics from those observed on the multipartite network. Transmission dynamics taking place on projected networks consistently overestimate the number of infected agents over the course of time. These results are less striking than those obtained for the physicians' network, but still consistent.

\subsection*{Factors controlling stability of unipartite projections}

The movie producers' network and physicians' network differ significantly in terms of the number of agents, the team sizes, the aggregation time scale, and the number of teams per aggregation cycle. Thus, we investigate which of these factors can explain the ability of projected networks to make better predictions. Additionally, we study a simple bipartite network model with community structure~\cite{Guimera2007,Colizza2006} in order to account for the possibility of modular structure in real-world multipartite networks (see Methods). 

\begin{figure*}[t]
\centerline{\includegraphics[width=0.8\textwidth]{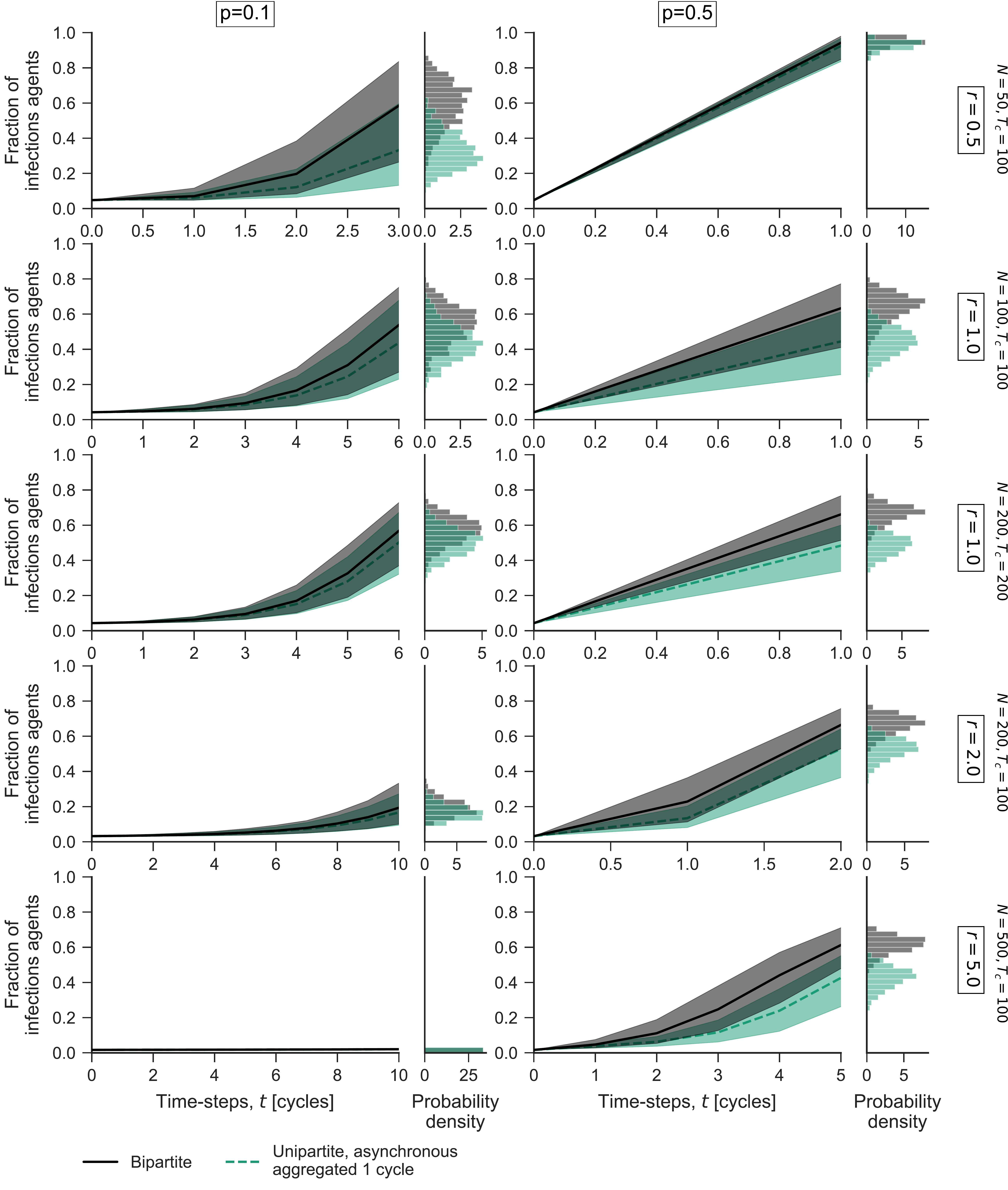}}
\caption{\textbf{Transmission dynamics on bipartite networks and their unipartite projections.}  Average and spread of transmission dynamics for the GC model using infection probabilities $p=0.1$ (left column) and $p=0.5$ (right column) and dose $d=0.5$. Note that as the ratio agents and teams (i.e., $r\equiv\frac{N}{T_c}$) increases, the difference in the dynamics becomes less pronounced.}
\label{fig:fig3}
\end{figure*}

We find that, as the degree of temporal coarse graining increases, the transmission dynamics of unipartite projected networks become more similar to those of temporal bipartite networks (Fig.~\ref{fig:fig3}). This is to be expected because the smaller the temporal graining is, the more the projected networks become an accurate snapshot of the bipartite networks. 

We also find that the number of agent $N$ increases whereas the number of teams per aggregation cycle $T_c$ remain fixed and the difference between the transmission dynamics of the bipartite networks and their projected networks also increases (Fig.~\ref{fig:fig3}). 

Our analysis, thus, reveals that the determining factor is the ratio of the number of agents to the number of teams per aggregation cycle. It is important to note, however, that the trajectory of the transmission dynamics for the projected and the bipartite networks have distinct functional forms. Thus, even the temporal fine-grained networks used in the context of temporal networks to obtain static snapshots~\cite{Holme2005} would not be sufficient to generate an accurate conceptualization of temporal multipartite networks that include a class of nodes with discontinuous dynamics. 

\section*{Building realistic multipartite networks}
Since simulation on unipartite projections cannot, for many conditions, replicate the transmission dynamics observed on multipartite networks, we ask whether there are suitable null models such that the dynamics on the generated networks better replicate the dynamics on real temporal multipartite networks. Specifically, we aim to identify the crucial information needed for creating adequate synthetic multipartite networks and we focus here on the physicians' networks because they show the largest dynamical differences between projected and multipartite networks.

In building our multipartite network models, we focus on three types of aggregate information: the number of agents in each agent class of nodes; distribution of the number of teams that an agent participates in; and distribution of gap durations between consecutive participations in a team by an agent. We construct three null models by using increasing amounts of information about the real network [Fig.~\ref{fig:fig4}(a)]. The first generative model, denoted \textbf{N}, makes use only of the number of agents $N$ in the system. The second model, denoted \textbf{NT}, adds also information on the distribution of the number of teams $T$ in which agents participate [Fig.~\ref{fig:fig4}(b) top]. The third model, denoted \textbf{NTG}, adds information on the distribution of gap $G$ durations between participation in consecutive teams as well [Fig.~\ref{fig:fig4}(b) bottom]. 

Figure~\ref{fig:fig4}(c) shows the level of information that is used in each model and the steps for generating a synthetic multipartite network. We show the distributions of the number of teams and gap durations for the three types of synthetic networks and the real networks in Fig.~S8 of the Supplemental Material~\cite{SM}.

\begin{figure*}[!b]
\centering 
\includegraphics[width=0.95\textwidth]{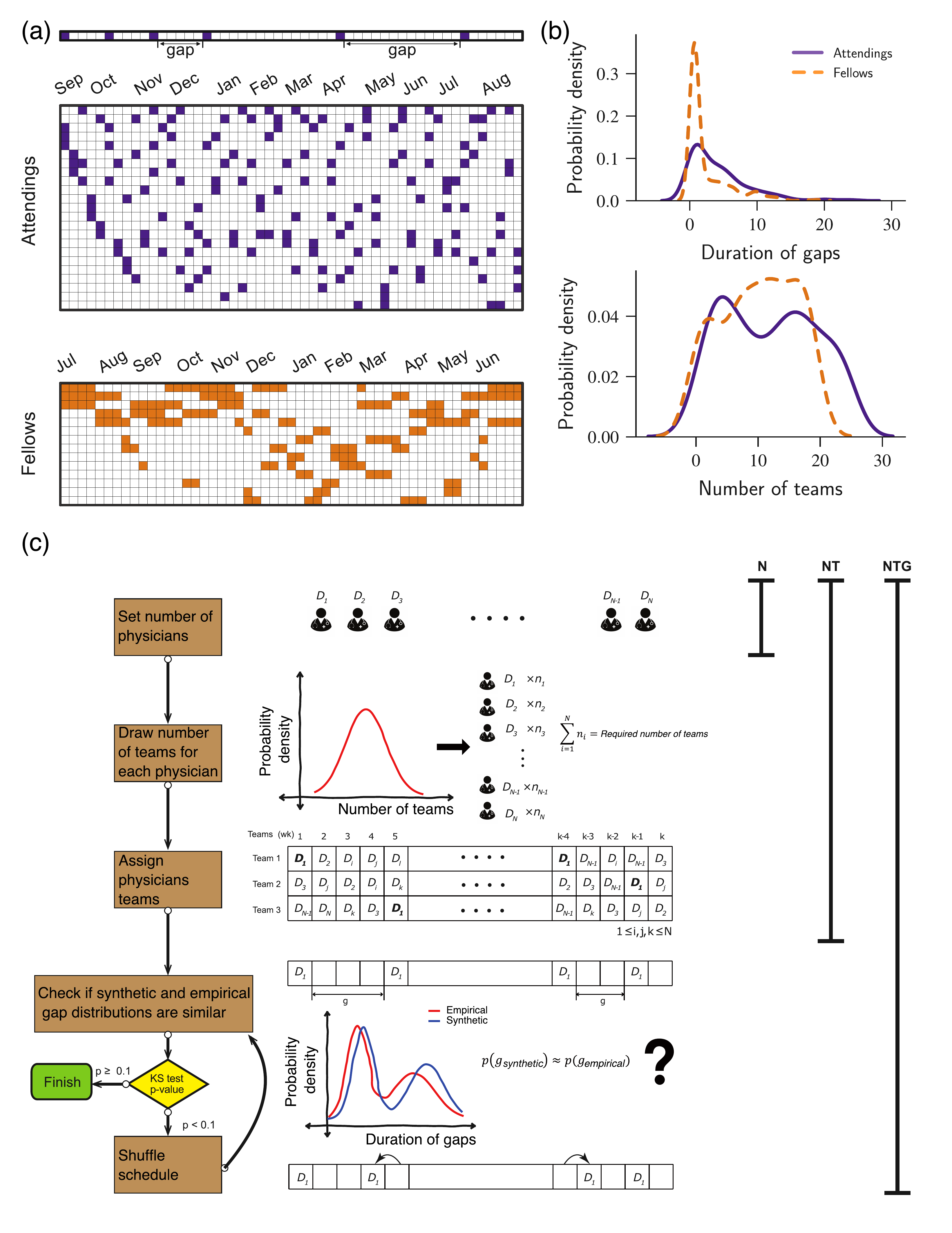}
\end{figure*}
\begin{figure*}[!t]
\caption{\textbf{Generation of synthetic multipartite networks.} (a) One year schedule for attendings and fellows in the MICU. Each row displays the entire year schedule for a given physician. Each square represents one week and filled squares indicate participation in a coverage team by a given physician. Consecutive white squares show gap durations between participation in coverage teams for each physician. (b) Distribution of the number of coverage teams a physician is part of during a year and distribution of the duration of gaps between participation in consecutive coverage teams. We first note that whereas coverage teams are reportedly assigned randomly by the hospital based on the number of teams that each physician has to be assigned to, we find that the distribution of the number of teams in which physicians participate does not obey a Poisson distribution~\cite{Masuda2017}. Therefore, physicians' team assignments cannot be purely random. The distributions of gap durations are also not exponential, which is the expected distribution for inter-event times that occur at a constant rate~\cite{Keeling2005}. These two findings suggest that there are other constraints that govern how physicians are assigned to teams. (c) Algorithm for generating synthetic multipartite networks according to levels of information incorporated. We create $N$ physician nodes as in the multipartite network we are trying to replicate (the \textbf{N} model). We can then either draw the number of coverage teams assigned to a physician from a Poisson distribution or from the distribution of the number of teams in the multipartite network without replacement (the \textbf{NT} model). Note that the total number of a team must be equal to the number in the multipartite network. We assign physicians' team participation at random. If we are including information about the distribution of gap durations, then we calculate the gap duration distribution for the synthetic network and compare it to that of the multipartite network. The teams are adjusted to match the gap duration distribution until the difference between the synthetic network and the multipartite network is not statistically significant (the \textbf{NTG} model).}
\label{fig:fig4}
\end{figure*}

\subsection*{Quantification of the similarity of two ensembles of dynamical trajectories}
Next, we address how to quantify the differences in the outcomes between the transmission dynamics on the synthetic networks and that on the real networks. To this end, we define a new metric, the dynamic overlap of two ensembles of dynamical trajectories, which we denote as $\Omega$. We focus on the GC model since the SI model is recovered for $D=d$. We compare both the final number of infectious nodes as well as the similarity of their trajectories when using the same parameters, on either synthetic or empirical networks. Quantifying such a comparison is not trivial and different methods have been introduced to compare the outcomes of transmission dynamics in networks~\cite{deArruda2018}, however, there has been less focus on comparing the transmission trajectories over the entire period. Inspired by the weak Fr\'echet distance~\cite{Alt1995}, we first define the distance between two dynamic trajectories as the area between the curves for an observation period $T$
\begin{equation}
A_j^i ( {\cal N}_1, {\cal N}_2 ) \equiv \int_0^T ~ | ~ f_i( t | {\cal N}_1 ) - f_j( t | {\cal N}_2 ) ~ | ~  dt\,,
\end{equation}
where $f_i( t | {\cal N}_1 )$ is the $i^{th}$ realization of the dynamics on network ${\cal N}_1$. We control for the potentially different number of agents in each network by scaling the trajectories by the number of agents in the network. We demonstrate how individual realizations of the dynamics on the same network can yield quite distinct trajectories in the top panel of Fig.~\ref{fig:fig5}A(a). Thus, our aim must be to quantify whether a particular trajectory on a network falls within the expected range for trajectories obtained for simulations of the dynamics on a set of reference networks, ${\cal R} = \{ {\cal R}_1, {\cal R}_2, \dots \}$. To this end, we define the distance of trajectory $i$ to the mean trajectory $R$ on a set of reference networks as
\begin{equation}
A_R^i( {\cal N}_k,{\cal R}) \equiv \int_0^T ~ |~ f_i( t | {\cal N}_k ) - \overline{f_R( t | {\cal R} )} ~ | ~ dt\,,
\end{equation}
where  $\overline{f_R( t | {\cal R} )}$ represents an average over realizations of the simulations and over the set of networks in ${\cal R}$ [middle panel, Fig.~\ref{fig:fig5}(a)]. By calculating $A_R^i( {\cal N}_k,{\cal R})$ for the $i^{th}$ trajectory, we can estimate the percentiles for the distances to the mean trajectory, $P_{2.5} \{A_R^i( {\cal N}_k,{\cal R)} \}$ and $P_{97.5} \{A_R^i( {\cal N}_k,{\cal R)} \} $, that  encompass 95\% of the dynamical trajectories generated on the set of reference networks [bottom panel, Fig.~\ref{fig:fig5}(a)]. We define the dynamic overlap $\omega$ for a given network ${\cal N}_k$ as
\begin{equation}
\omega ( {\cal N}_k ) \equiv \int_{P_{2.5}}^{P_{97.5}}  ~ A_R ~ p( A_R |  {\cal N}_k ) ~  dA_R\,,
\end{equation}
where $ p( A_R |  {\cal N}_k ) $ is the probability of observing a specific value of $A_R^i( {\cal N}_k,{\cal R}) $. The dynamic overlap for a set of multipartite networks is $\Omega \equiv P_{50}\{\omega({\cal N}_k)\}\,$, which takes values in $[0, 1]$. In practice, we define ${\cal R}$ for multipartite networks as the set of networks for 2012, 2013, and 2015 and use the real network for 2014 as a control for estimating the similarity of the trajectories for different parameter values. We chose ${\cal R}_{2014}$ as the control because its number of physicians takes a value close to the mean for the four networks. We then compare the trajectories of the GC model for the control network with results obtained for synthetic multipartite network generated using information from ${\cal R}_{2014}$ [Fig.~\ref{fig:fig5}(b)] .

\begin{figure*}[!b]
\centering 
\includegraphics[width=0.95\textwidth]{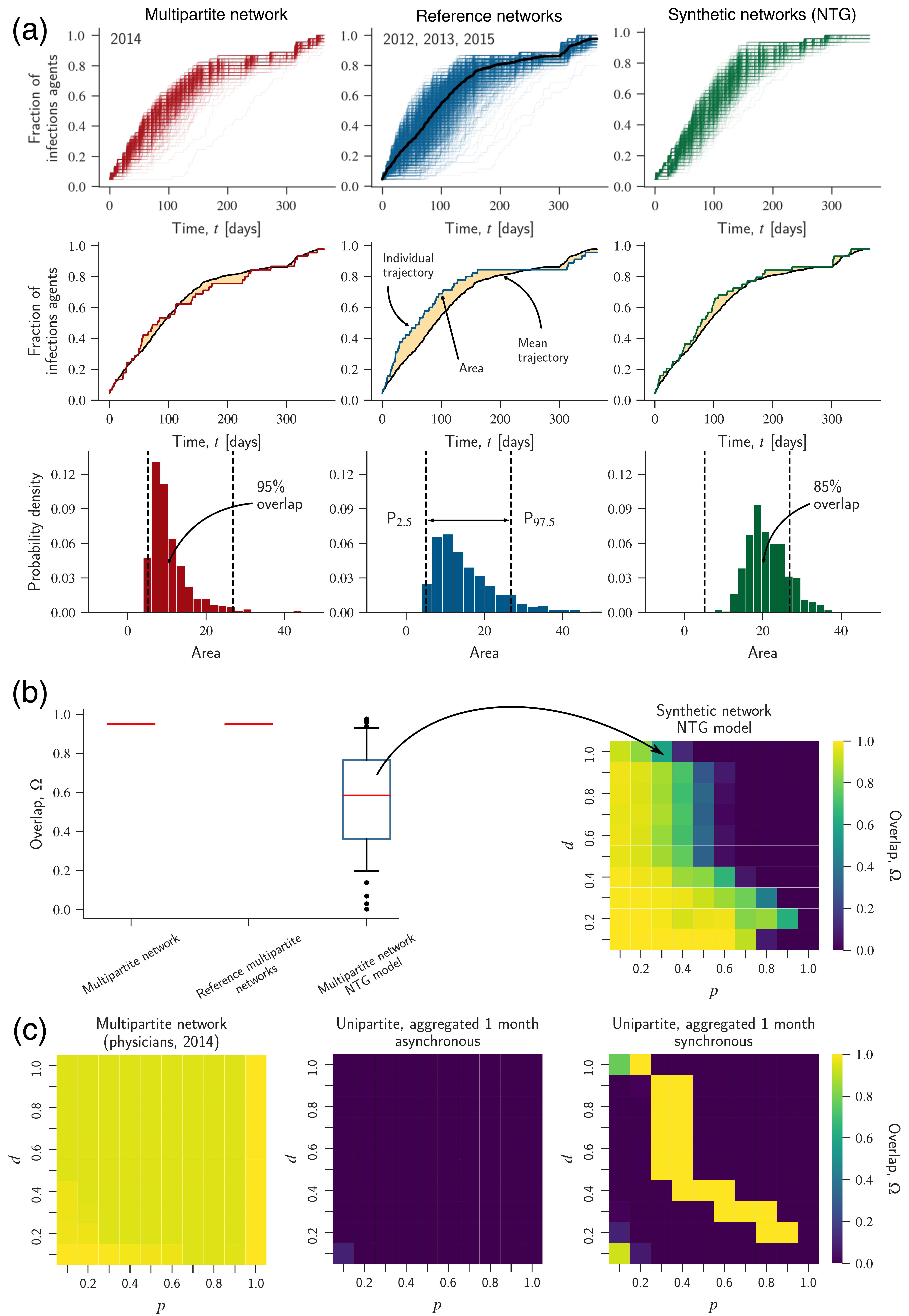}
\end{figure*}
\begin{figure*}[t!]
\caption{\textbf{Systematic quantification of the similarity of transmission dynamics on different networks.} (a) The top panel shows the simulation of 1,000 independent stochastic realizations of the GC model dynamics for each network: the 2014 control temporal multipartite network (red), the reference set of multipartite networks for 2012, 2013, and 2015 (blue), and 25 synthetic networks (green). The middle panel shows the areas between the mean of realizations from the reference networks (reference curve, i.e., thick black line in the top panel) and each realization from the reference networks. The bottom panel shows the area distribution $ p( A_R |  {\cal N}_k ) $ where we identify the 95\% confidence boundaries for the observed overlapping. The percentages indicate the overlapping between the areas from the simulations compared with those from the reference set. (b) The panel shows the box-plot of the area overlapping for the control multipartite network, the reference set, and the synthetic network for a specific set of parameters, as indicated in the figure. We calculate $\omega({\cal N}_k)$ for each set of networks in multipartite networks and synthetic networks. Because there are 25 synthetic networks, we define $\Omega$ as the median of $\omega({\cal N}_k)$ for the 25 networks.  (c) The heat maps show the dynamic range overlap $\Omega$ for the entire parameter space for different networks. Each cell shows the $\Omega$ value of the GC model dynamics compared with those obtained on the reference set ${\cal R}$ for the multipartite network (left), for projected networks using synchronous (middle), and asynchronous (right) update. }
\label{fig:fig5}
\end{figure*}

We show $\Omega$ for the control multipartite network ${\cal R}_{2014}$ --- multipartite network from the hospital for the year 2014 in Fig.~\ref{fig:fig5}(c). Similar results are obtained for $\Omega$ when using ${\cal R}_{2012}, {\cal R}_{2013}$ and ${\cal R}_{2015}$ as control networks (see Figs.~S9 to S11 in the Supplemental Material~\cite{SM}). Even though ${\cal R}_{2014}$ is not structurally different from the networks in ${\cal R}$, we still found that $\Omega$ is close to zero when $p$ is large. This is due to the fact that the stochasticity of the transmission dynamics becomes negligible when almost all interactions lead to successful transmission. Under these conditions, even small differences in the network lead to statistically significant differences in the dynamics (Fig.~S1 in the Supplemental Material~\cite{SM}).

In contrast, for the projected networks, $\Omega$ is generally close to zero for nearly all regions of parameter space [Fig.~\ref{fig:fig5}(c), and Figs.~S12 to S15 in the Supplemental Material~\cite{SM}]. The exceptions are those parameter values for which the variance in the trajectories for the multipartite networks is very large. Because the trajectory diversity for the dynamics on the projected networks are usually small (Fig.~\ref{fig:fig2}), it enables all the trajectories on the projected networks to fall within the 95\% confidence interval of the dynamics on the multipartite networks (see the case of $p = 0.1$ in Fig.~\ref{fig:fig2} and Figs.~S1 to S6 in Supplemental Material~\cite{SM}). This interpretation is supported by the fact that when we take the time-aggregated unipartite networks as the reference set and the multipartite networks as the ones being compared, then $\Omega$ drops to zero (Figs.~S16 and S17 in the Supplemental Material~\cite{SM}). If the transmission dynamics were truly similar, $\Omega$ would take similar values even when ${\cal R}$ and $\{{\cal N}_k\}$ are switched so that $\{{\cal N}_k\}$ are used as reference networks, as indeed is the case for synthetic multipartite networks (Fig.~S18 in the Supplemental Material~\cite{SM}). 

\begin{figure*}[!t]
\centerline{\includegraphics[width=0.793\textwidth]{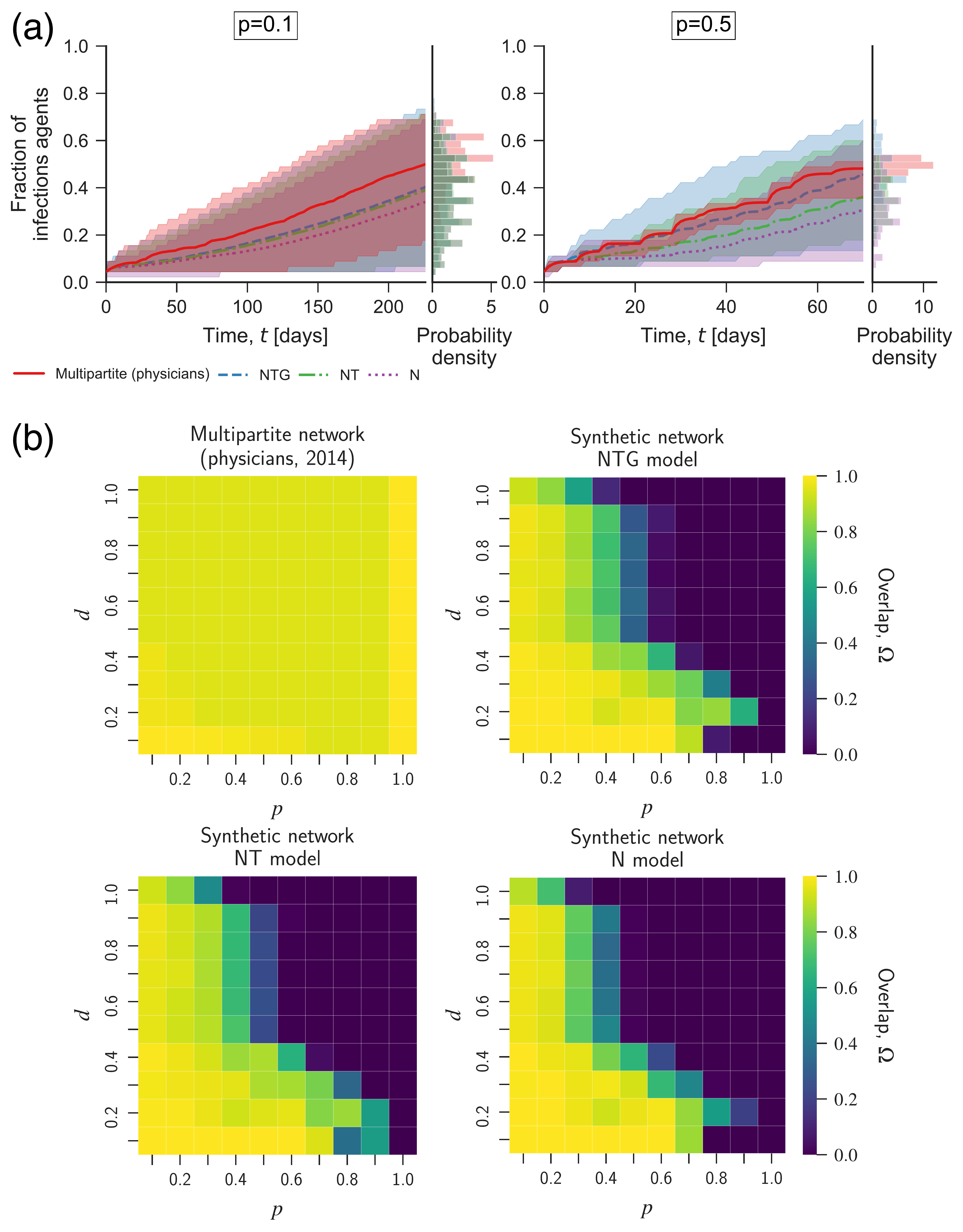}}
\caption{{\bf Model dynamics on multipartite networks and on their synthetic networks.} (a) Mean fraction of infectious agents over time (lines) and 95\% confidence interval for the dynamics (shaded regions) on the multipartite and synthetic multipartite networks using SI model with infection probabilities $p=0.1$ and $p=0.5$.  The histograms show the final fraction of infectious agents for each network.  (b) The heat maps show the dynamic range overlap $\Omega$ for the entire parameter space for different networks. Each cell shows the $\Omega$ value of the GC model dynamics compared with those obtained on the reference set ${\cal R}$ for the multipartite network (top left) and the three synthetic multipartite model networks.
}
\label{fig:fig6}
\end{figure*}

Thus, we find that as we increase the real-world information built into the null models, the more closely the infection trajectories tend to mirror those observed for the control network [Fig.~\ref{fig:fig6}(a)]. In fact, calculating the overlap $\Omega$ of the transmission dynamics in synthetic multipartite networks with the reference set of network ${\cal R}$ for different values of parameter of the GC model, there is a strong agreement between the transmission trajectories and the control network for a wide range of parameters [Fig.~\ref{fig:fig6}(b), and Figs.~S9 to S11 in the Supplemental Material~\cite{SM} for comparisons using other years than 2014 as control network]. Thus, even simplistic synthetic networks can produce $\Omega$ values that are quite similar to the values obtained for the control network and can capture real dynamic much better than the projected networks (Fig.~S19 in the Supplemental Material~\cite{SM}). 

Furthermore, to verify that this model is generalizable to other cases, we generate synthetic multipartite networks using the characteristics of the producers' network. We randomly selected 5\% of the producers active in the first year as initial infectious agents. We found that the transmission dynamics on even the simplest synthetic network are much more similar to the transmission dynamics on the original multipartite network than on projected networks which is consistent with the results from the physicians' network. However, we observed that the infection rates are not as similar to the real multipartite network compared to the physicians' case, although the overall shape of the infection curves is quite similar (Fig.~S20 of the Supplemental Material~\cite{SM}). This difference may be due to the fact that unlike for the physicians' networks, producer networks are more heterogeneous. An alternative or even complementary explanation is that physician and producer teams have different formation mechanisms. In the physicians' networks, one attending and one fellow form a team, whereas teams vary in size and composition for the producers' network (Figs.~S21 and S22 in the Supplemental Material~\cite{SM}). 

\section*{Discussions}

Transmission dynamics on systems best described by temporal multipartite networks show much greater variability across realization than one is led to believe from simulating dynamics on time-aggregated unipartite projection networks. This difference means that studies conducted on projected networks are inadequate both for obtaining estimates of expected outcomes and deciding whether an observed outcome is consistent with a given model. We show that the ratio of the number of agents to the number of teams per aggregation cycle has a significant impact on the resulting differences of simulating the transmission dynamics on multipartite networks versus their unipartite projections. This means that there is the need to accomplish a trade-off between decreasing the number of teams per aggregation cycle whereas maintaining a small ratio of the number of agents to the number of teams per aggregation cycle.

Additionally, we show that simulations on very simplistic multipartite synthetic networks can capture real dynamics much better than simulations on projected networks. Our results inform about the types of information to focus on when there are limited resources for collecting data on complex multipartite networks.  

Finally, because of the growing body of information-rich networks datasets~\cite{datasets}, our results open new venues to explore a variety of systems that can be represented as temporal multipartite networks, including very large networks with different topological heterogeneity that could result in new behaviors that cannot be unveiled by simple time-aggregated unipartite projected networks.

\section*{Methods}
\subsection*{Transmission models}
We study the transmission dynamics as a contagion process diffusing either on the actual temporal multipartite network, or on different projections obtained for different levels of temporal aggregation, or on synthetic multipartite networks. We consider two contagion models: the SI model and the general GC model. 

{\it SI model:} This model agents can be in one of two states, susceptible ($S$) and infectious ($I$), which we represent as 0 and 1, respectively. At each event or time step $t \geq 1$, a susceptible node in contact with an infectious node becomes infected with probability $p$. 

{\it GC model:} This model interpolates between two classes of transmission mechanisms: threshold~\cite{Granovetter1978} and infection~\cite{Anderson1992}. When a susceptible agent is exposed to infectious agents, they receive an infection dose $d$ with probability $p$. Susceptible agents keep in memory the infection doses received over the previous $\tau$ time steps. Susceptible agents become infected when their cumulative remembered infection doses exceed a threshold, $D$ (see algorithm 1 in the Supplemental Material~\cite{SM}). For simplicity, we restrict our attention to the case where $\tau = \infty$. For both transmission models, each simulation is stochastic and starts with two infectious nodes chosen randomly from a fully susceptible population. Note that the SI model is captured by the case where $d=D$ in the GC model.

\subsection*{Simulation of dynamics}
We systematically study the transmission dynamics for $p \in \{0.1, ... , 1.0\}$ (SI) and for 100 pairs of values $p \in \{0.1, ... , 1.0\} \times d \in \{0.1, ... , 1.0\}$ (GC) and for four different time scales of aggregation in the unipartite projection. For each set of parameter values and each network, we simulate 1,000 independent realizations. 

{\it Unipartite network dynamics:} In order to assure the robustness of our results for unipartite projected networks, we consider two methods to update the state of the agents --- asynchronous update using a random walk procedure~\cite{Tang2009,Aggarwal2011}, and synchronous update~\cite{Granovetter1978,Shakarian2015,Chen2010,Valente1996}. 

In the asynchronous update, at the start of transmission, $t=0$, we randomly select two nodes as initially infected nodes $n_1, n_2$. At each time step, a node $n_1$ is selected and one of its neighbors is randomly choose to calculate the transmission dynamic. Then we repeat the previous step $t$ times according to the time interval the network was projected and the number edges in the network. The probability of choosing the neighbor and the size of the influence two active nodes have on each other is proportional to the edge weights. At each step only the chosen nodes interact (see algorithm 2 in the Supplemental Material~\cite{SM}). When one node is in the susceptible state and the other is in the infected state, the susceptible node receives a dose $d$ with probability $p$ for the GC model or is infected with probability $p$ for the SI model (see algorithm 1 in the Supplemental Material~\cite{SM}). If both nodes are susceptible or infected (when nodes are in the same state), no exchange occurs. Asynchronous update behaves as if it has temporal properties because the edges are selectively active for a certain duration.

In the synchronous update, at time step $t$, the transmission of infection can happen across all connected nodes, and all nodes update their state simultaneously (see algorithm 3 in the Supplemental Material~\cite{SM}). This updating strategy is inspired by a modified linear threshold model, where a susceptible node is influenced by all of its infected neighbors at each time step~\cite{Valente1996,Granovetter1978,Shakarian2015,Chen2010}. The influence a node exerts on another depends on the edge weights. In the synchronous update, it is assumed that edges always existed during the interval of the simulation, therefore, in a network of finite size and non-zero transmission rate, all nodes will eventually become infected~\cite{Gross2006}.

{\it Multipartite network dynamics:} For the multipartite networks, the transmission can progress through the active members of the coverage team; physicians who are part of a team at each time point interact and influence each other. When one node is in the susceptible state and the other is in the infected state in an active team, the susceptible node receives a dose $d$ with probability $p$ for the GC model or is infected with probability $p$ for the SI model. However, because a certain team can be active during different periods, for each day, the infection process is repeated until the team is no longer active (see algorithm 4 in the Supplemental Material~\cite{SM}). 

For the producers' multipartite network, because we do not have data on exact production periods, we randomized the order of the movie production during each year to simulate the transmission dynamics. The number of movies produced per year stays approximately constant (Fig.~S22 in the Supplemental Material~\cite{SM}), therefore we can assume the gap between produced movies remains approximately constant. 
\subsection*{Generating the synthetic bipartite network with a modular structure}
We use the generative model for bipartite networks introduced by Guimer\`a \textit{et al.}~\cite{Guimera2007}, because it yields an ensemble of random bipartite networks with a prescribed modular structure. To create a network instance, we first dived $N$ agents equally into five groups. We then sequentially create $N_T$ teams. Each time, we first select at random a primary group. Then, with probability $p_p$ (team homogeneity) we select an agent at random from the primary group. With probability $1-p_p$, we select an agent at random from one of the four non-primary groups.  Thus, we use the generated bipartite networks to systematically investigate the properties of the bipartite networks that affect the dynamics of the projected networks.

\subsection*{Acknowledgements}
{We would like to thank M. Gerlach and J. Poncela-Casasnovas for their thoughtful consideration and feedback on an early version of this work. We would like to acknowledge financial support from the Department of Defense Army Research Office, Grant W911NF-14-1-0259 and John, and Leslie McQuown.  The funder had no role in study design, data collection, and analysis, decision to publish or preparation of the paper. 
}

\subsection*{Author contributions}
H.A.L and L.G.A.A. contributed equally to this work. H.A.L,  L.G.A.A., and L.A.N.A. conceived and designed the study. H.A.L and  L.G.A.A. performed the numerical simulations and statistical analysis, created the figures, and wrote the first draft of the paper.  H.A.L, L.G.A.A., and L.A.N.A. wrote, read and approved the final version of the paper.

\subsection*{Conflicts of interest}
The authors declare no competing interests. 


%

\end{document}